# New Correlations between Viscosity and Surface Tension for Saturated Normal Fluids

**Mengmeng Zheng [1], Jianxiang Tian [1, 2, 4], Ángel Mulero[3]**

[1]Shandong Provincial Key Laboratory of Laser Polarization and Information Technology

Department of Physics, Qufu Normal University, Qufu, 273165, P. R. China

[2]Department of Physics, Dalian University of Technology, Dalian, 116024, P. R. China

[3]Department of Applied Physics, University of Extremadura, Badajoz 06006, Spain

[4]Corresponding author, E-mail address: jxtian@dlut.edu.cn

## Abstract

*New correlations between viscosity and surface tension are proposed and checked for saturated normal fluids. The proposed correlations contain three or four adjustable coefficients for every fluid. They were obtained by fitting 200 data points, ranging from the triple point to a point very near to the critical one. Forty substances were considered, including simple fluids (such as rare gases), simple hydrocarbons, refrigerants, and some other substances such as carbon dioxide and water. Two correlation models with three adjustable coefficients were checked, and the results showed that the one based on the modified Pelofsky expression gives the better overall results. A new 4-coefficient correlation is then proposed which clearly improves the results, giving the lowest overall deviations for 32 out of the 40 substances considered and absolute average deviations below 10% for all of them.*

## Introduction

Surface tension and viscosity are two properties of fluids which are different in nature but whose values need to be known for a wide variety of industrial and physicochemical processes. Surface tension affects important stages in processes such as catalysis, adsorption, distillation, and extraction, and viscosity is important in processes involving a flow of fluids, such as the use of lubricants. The two properties have been extensively studied for normal fluids, and this interest continues (see [1-5] for instance).

Fluid viscosity, $\eta$, can be measured with high precision, and the resulting data and its temperature dependence are used as essential properties for the accurate determination of molecular information such as the pair interaction potential function [2]. Low-temperature viscosity correlations usually assume that $\ln \eta$ is a linear function of reciprocal absolute temperature [1]. In the region from about $T_r = T/T_c = 0.7$ K (where $T_c$ is the temperature at the critical point) to near the critical point, there are many complex equations available that permit one to express the temperature dependence of viscosity. Examples are the Sastri [1], Orrick and Erbar [6-7], and Vogel-Fulcher-Tamman (VFT) equations [8-9]. Of these equations, the VFT one has proven to be the most accurate, and has been widely used for research into ionic liquids [9-10], hydrogen-bonded fluids [10], and even the early Earth's magma ocean [11].

Surface tension, $\sigma$, is also related to the intermolecular interaction potential energy and the liquid interfacial microstructure [1, 12-15]. It can also be measured with high accuracy at low and moderate temperatures and pressures. Nevertheless, at high temperatures and high pressures, computer simulations are usually required [4, 16].

Experimental results show that surface tension is a linear function of temperature $T$ for values of $T_r$ between 0.4 and 0.7 [1]. At higher temperatures, the surface tension is usually expressed [1, 17-22] as proportional to one or more terms of the form $(1-T_r)^n$, where $n$ is a fixed constant or substance-dependent coefficient.

For some fluids, one of these two properties may be more easily measured than the other for certain temperature ranges. Moreover, as indicated previously, both properties are related to the microscopic structure and intermolecular forces of fluids. It is therefore interesting to try to establish some relationship between them. Such a relationship could also be used to test the validity of the measured data, since any deviations may be due to experimental error [23]. Indeed, since both properties are related to the intermolecular potential energy, one might expect there to be some theoretical correlation between the two, although no such link has yet been established.

In 1966, Pelofsky [24] proposed an empirical relationship between the natural logarithm of surface tension and the inverse of viscosity (usually termed the fluidity). Two adjustable coefficients are needed whose values may depend on the temperature range being considered. This correlation was later modified by Schonhorn [25] who introduced a correction into the second term of the right-hand side of the expression to fulfill the requirement that, at the critical point, the surface tension goes to zero while the viscosity tends to a small constant value. This modification introduces new coefficients, and has not subsequently been used.

Queimada *et al.* [23] checked the use of the Pelofsky correlation for pure compounds and mixtures of n-alkanes, and found adequate results in all cases. The temperature ranges they considered were, however, fairly narrow, and indeed the authors themselves observed that near the critical point the results may be very inaccurate.

More recently, Ghatee *et al.* [26-27] applied the Pelofsky correlation to some ionic fluids. They found that it was necessary to modify it slightly by introducing an exponent into the viscosity term (we shall denote this hereafter as the modified Pelofsky, or MP, correlation). They initially treated this exponent as an adjustable coefficient, but then they found that its value could be fixed at 0.3 without any significant loss of accuracy for the fluids considered. The same correlation and fixed coefficients have very recently been used for the case of seven types of honey with different concentration and source types [28].

Both the original and the modified Pelofsky correlations have recently been studied by us for a set of 56 normal fluids [29]. We found that the performance and the accuracy of the Pelofsky expression in the calculation of the surface tension are very limited for the selected fluids and temperature ranges. In the case of the MP expression, which has one more adjustable coefficient, the results are clearly improved. Unfortunately, unlike the case of some ionic liquids, the exponent in the MP correlation did not take a fixed value.

In this present paper, we focus our attention on the calculation of the viscosity from the knowledge of the surface tension values for normal saturated liquids. For that, we consider the whole temperature range from the triple point to the critical point. We consider the MP correlation and propose two new ones. In Section 2, we describe the two new models and the previous MP one. In Section 3, we illustrate the results and discuss them. Finally, we give the conclusions.

## 1. Viscosity-surface tension correlations

Pelofsky proposed a relation between the surface tension and the viscosity as [24]

$$\ln \sigma = \ln A + \frac{B}{\eta} \tag{1}$$

where $A$ and $B$ are substance-dependent constants. According to Pelofsky [24], this empirical relation can be applied to both the organic and the inorganic phases of pure and mixed components. We have recently studied its accuracy for 56 fluids [29] by calculating the absolute average deviation (AAD) values for the prediction of the surface tension. We found that the AAD values are less than 2% only for four refrigerants and nonane. Moreover, AADs greater than 20% were found for water, oxygen, and deuterium oxide, for which compounds the P model is therefore clearly inadequate, at least for the wide temperature range considered. The previous results are improved when the MP expression, proposed by Ghatee et al. for ionic liquids [26], is used. This expression is as follows:

$$\ln \sigma = \ln C + D \left(\frac{1}{\eta}\right)^{\phi} \tag{2}$$

where $C$, $D$, and the exponent $\phi$ are substance-dependent coefficients. We have found [29] that this correlation improves the results with respect to the previous one, in part due to the presence of one more adjustable coefficient. With the use of the MP correlation, the surface tension data were reproduced with AADs below 2% for 13 out of the 56 fluids considered in [29], the poorest value being 7.3%. It was also found that the improvement with the MP correlation was very significant for 34 fluids.

As in the present paper we are interested in the calculation of the viscosity, the alternative form of Eq. (2) is used:

$$\left(\frac{1}{\eta}\right)^{\phi} = A_1 + B_1 \ln \sigma \tag{3}$$

One of the inconveniences of this expression is that it cannot be applied just at the critical point, where the surface tension is defined as zero and the natural logarithm is not defined. It is desirable, therefore, to introduce an alternative expression with a similar number of adjustable coefficients, but which does not include the logarithm of the surface tension.

It is well known that, for values of $T_r$ between 0.4 and 0.7, the surface tension-temperature relation can be represented by a linear equation as [1]:

$$\sigma = a + bT \tag{4}$$

with $a$ and $b$ being substance dependent constants. At higher temperatures, the following equation can be used as a first, but accurate approximation [1, 17-21] for a large variety of fluids:

$$\sigma = a_1(1-T_r)^n \tag{5}$$

The coefficient $a_1$ can be obtained as an adjustable coefficient or can be related to the critical properties [1, 18, 30], the acentric factor [18], or the Riedel parameter [30].

The most commonly used equation for describing the temperature dependence of viscosity is the Arrhenius equation [9]:

$$\eta = A_0 \exp(E_a / k_B T) \tag{6}$$

where $E_a$ is the activation energy for viscous flow and $k_B$ is the Boltzmann constant.

Another widely used expression is the Vogel-Fulcher-Tamman (VFT) equation [8-11]:

$$\eta = \eta_0 \exp[DT_0 /(T - T_0)] \tag{7}$$

where $\eta_0, D$, and $T_0$ are constants. The physical meaning of $T_0$ is the ideal glass transition temperature, which is the temperature of viscosity divergence. The VFT equation was used recently to describe the viscosity of fluid glycerol [8] and liquid $MgSiO_3$ in the Earth's mantle conditions [11].

Substituting Eq. (4) into Eqs. (6) or (7), one obtains the correlation between surface tension and viscosity as:

$$\ln \eta = A_2 + \frac{B_2}{\sigma + C_2} \tag{8}$$

This expression, which we shall call **ZTM3**, includes three adjustable coefficients, as also does the MP. A clear difference here is that one of the MP coefficients is an exponential whereas here only linear coefficients are used. Relating Eq. (5) to Eqs. (6) or (7), one obtains the expression:

$$\ln \eta = A_3 + \frac{B_3}{\sigma^{\frac{1}{n}} + C_3} \tag{9}$$

which contains four adjustable coefficients, one of them an exponent, *n,* and which we shall call the **ZTM4** correlation. In Eqs. (8-9), $A_i, B_i$, and $C_i$ are coefficients that are related to those of Eqs. (4-7). Moreover, they can be used just at the critical point by setting the surface tension equal to zero, thus providing an estimate of the value of the viscosity at this point.

## 2. Results and Discussion

As the main aim of the present paper was to study the relationship between two properties, it was important to adequately select the source of the data used to this end. We thus selected the NIST Web Book [30] because the data it offers are sufficiently accurate and are publicly and straightforwardly available. The data on the saturation curves are limited to a maximum of 201 data points. Since the surface tension is defined as zero at the critical point and the viscosity at the critical point is not given for several substances, we excluded this datum, so that the default number of data for each fluid was 200. Nevertheless, we found that for certain fluids the surface tension and viscosity data are not both available for some low or high temperature ranges. So we finally considered only those fluids for which the NIST Web Book [31] provides the values of the surface tension and viscosity over the whole temperature range from the triple point to very near the critical point. Forty fluids were selected, including simple fluids (such as argon and other rare gases), simple hydrocarbons, refrigerants, and some other substances such as carbon dioxide and water. These substances are listed in Table 1, in alphabetical order for three kinds of substances: refrigerants, hydrocarbons, and other common fluids. The data start at the temperature $T_0$, which is the triple point temperature, and finish at the temperature $T_f$, which is automatically selected by the software in the NIST Web Book for the final number of data to be 200. The small difference between $T_f$ and critical temperature, $T_c$, can be observed in Table 1.

In the particular case of R143a, the data given in the NIST Web Book for the surface tension at low temperatures are not adequate. Firstly, this is because they present a small maximum at low temperatures instead of monotonically increasing as the temperature decreases; and, secondly, because these data clearly disagree with those considered recently in Ref. [21]. For R143a we therefore used the correlation proposed in Ref. [21] for the surface tension, instead of the data given in the NIST Web Book.

The data for the surface tension and the viscosity were used to check the behaviour of the proposed models, ZTM3 and ZTM4, as well as the previous model, MP. During the fitting procedure, those coefficients that minimize the AAD values were chosen. They are given in Tables 3-5. To calculate the AAD, we first calculated the percentage deviation (PD) between the values for the viscosity obtained from the correlation by introducing the surface tension as input, $\eta(\sigma_i)$, and the data offered by NIST [31], $\eta_i$, as follows:

$$\mathrm{PD}_i = 100\left(\eta(\sigma_i) - \eta_i\right)/\eta_i, \quad i = 1, 2, \ldots, 200 \tag{10}$$

A positive $\mathrm{PD}_i$ value means that the model overestimates the accepted datum, whereas a negative $\mathrm{PD}_i$ value means that the model underestimates it. Then we calculated the average absolute percentage deviation for every fluid:

$$\mathrm{AAD} = \frac{\sum_{i=1}^{N} |\mathrm{PD}_i|}{N} \quad (\%) \tag{11}$$

It has to be borne in mind that, since AAD is a percentage, it is influenced by the high individual PD values that can be found when the viscosity takes very low values (near to zero), which occurs at the highest temperatures, *i.e.* near the critical point temperature. This means that near the critical point, the absolute deviations are low, but the relative PD can take very high values and this has a clear influence on the final AAD value.

The AAD values obtained for the three correlations analysed are given in Table 1, and in Table 2 the results are summarized by showing the number of fluids for which each correlation gives an AAD value lower or higher than a given quantity. Tables 3 to 5 list the coefficients obtained for each correlation model. Figures for each substance with their viscosity values from the data and the three models, and the corresponding PDs, are available as Supplementary Material.

As can be seen in Table 1, we find that for 4 out of the 40 substances considered the MP correlation gives very poor results, with AAD values greater than 10%. Water is a clear example, as is shown in Fig. 1, which plots the PDs. One observes that the MP model reproduces the viscosity data well near the critical point where both the surface tension and the viscosity values are low, but not near the triple point. Although the other two models are not adequate near the critical point (the PD values increase to around 30-40%), they give better overall results.

The MP model gives AADs lower than 5% for 26 substances. Indeed, for most substances it gives better results than the ZTM3 correlation, and for 8 substances (see Table 1) it gives even better results than the ZTM4 model which includes one more adjustable coefficient. A clear example is dodecane, Fig. 2. For this substance, the MP model perfectly correlates the data over the whole temperature range, whereas the other models are accurate neither near the triple point (right part of the figure) where the absolute deviations are great, nor near the critical point (left part of the figure) where the relative deviations are high.

Unfortunately, the new model proposed here, ZTM3, gives worse overall results than the MP one. In most cases, it cannot reproduce the data near the critical point, so that high PDs are obtained in this region (see a clear example in Fig. 1 for water, as well as in Figs. 4-5, below). Therefore, it is desirable to know the performance and accuracy of the ZTM4 model, which includes one more adjustable coefficient. As can be seen in Tables 1 and 2, the ZTM4 correlation, Eq. (9), gives AAD values below 10% for all the fluids considered. The worst results are obtained for six fluids for which the AAD values range from 5% to 10%. For three of these fluids (dodecane, nonane, and pentane) we found that the MP correlation gives better results (a clear example is shown in Fig. 2). For the other 3 substances (R13, isobutene, and propane) none of the correlation models gives an AAD below 5%. We found that these three substances have in common that the viscosity data are almost constant and very near to zero at high temperatures, whereas they strongly increase at low temperatures. An example is shown in Figs. 3 and 4 for propane. As can be seen there is a clear difference between the behaviour of the data for the viscosity at high (left part of Fig. 3) and low temperatures (right part of Fig. 3). None of the models lead to an adequate fit at both extremes, and the PDs are too high (Fig. 4).

The best results (AAD < 1%) for the ZTM4 model are obtained for four common substances and one refrigerant, RC318. As can be seen in Fig. 5, the improvement with respect to the other correlations for these fluids is significant over the whole temperature range.

As noted above, a clear difference in the use of both the ZTM3 and the ZTM4 models with respect to the MP one is that the first two can be used just at the critical point by setting the surface tension equal to zero. This would allow one to estimate values of the viscosity at the critical point.

As can be seen in Figs. 1, 4, and 5, the ZTM3 expression is far from giving good results near the critical point (surface tension near to zero). Estimated values from ZTM4 and the percentage deviations for 21 substances, those for which the NIST Web Book gives values for the viscosity just at the critical point, are listed in Table 6. As can be seen, the results are very irregular. For 9 out of the 21 substances we obtain absolute values for PD < 10%. The worst value is obtained for hydrogen sulfide, despite the ZTM4 model giving an AAD value of 0.1% for the whole temperature range. It is clear that it is very difficult to obtain low PD values due to the fact that the viscosity is almost zero at the critical point, and hence the relative deviations are high even though the absolute deviations are low in some cases.

## 3. Conclusions

Three models for the correlation of the viscosity versus the surface tension have been checked for forty fluids of different kinds. Data from the NIST Web Book [31] were considered as referents, except in the case of R143a, for which Ref. [21] was considered as more appropriated. The results for the viscosity data were tested by obtaining percentage deviations for every datum and the absolute average deviation for each fluid. Figures for every fluid are available as Supplementary Material.

By comparing the ZTM3 and MP models, we found that the MP one improves the results for most of the fluids considered, giving the better overall results for 8 of them. In these cases, the MP model, with only three adjustable parameters, is therefore better than the ZTM4 one, which has four adjustable parameters. However, the MP model is not accurate (AAD > 10%) for four fluids. Moreover, we have shown that the ZTM3 model cannot be used with accuracy at the critical point.

The ZTM4 model is clearly the one giving the best results, with AAD <10% for all the fluids considered, being lower than 1% for five of them. Only for six fluids were AAD values greater than 5% found. In the particular cases of R13, isobutene, and propane, none of the models considered here can give AAD values lower than 5%. However, the improvement using ZTM4 is clear for most of the fluids (see Table 1). Finally, we found that ZTM4 could be used with accuracy just at the critical point only for 9 out of the 21 fluids for which the NIST Web Book gives this value.

Although there is some room for improvement by developing new correlation models connecting the viscosity and the surface tension of fluids, the ZTM4 model proposed here is clearly adequate, and is based on the study of the temperature behaviour of both properties over wide temperature ranges. The MP model is also a good alternative for some kinds of behaviour. Both models will be considered in the future to study this correlation for other kinds of fluids for which only more limited data are available.

## Acknowledgements

The National Natural Science Foundation of China under Grant No. 11274200, the Natural Science Foundation of Shandong Province under Grant No. ZR2011AM017, and the foundations of QFNU and DUT have supported this work (M.Z. and J.T.). It was also partially supported by the "Gobierno de Extremadura" and the European Union (FEDER) through project GR10045 (A.M.)

**FIGURE CAPTIONS**

**Figure 1**. Percentage deviations for the calculation of the viscosity for water from three equations. Dotted line: MP; dashed line: ZTM3; continuous line: ZTM4.

**Figure 2.** Viscosity versus surface tension for dodecane. Circles: NIST data; dotted line: MP; dashed line: ZTM3; continuous line: ZTM4.

**Figure 3.** Viscosity versus surface tension for propane. Circles: NIST data; dotted line: MP; dashed line: ZTM3; continuous line: ZTM4.

**Figure 4.** Percentage deviations in the calculation of viscosity for propane from three equations. Dotted line: MP; dashed line: ZTM3; continuous line: ZTM4.

**Figure 5**. Percentage deviations in the calculation of viscosity for argon from three equations. Dotted line: MP; dashed line: ZTM3; continuous line: ZTM4.

**Table 1.** Average absolute deviation (AAD) values for the viscosity of fluids obtained by using the MP, ZTM3, and ZTM4 correlations, which are Eqs. (3), (8), and (9), respectively. The initial and final temperatures, $T_0$ and $T_f$, as well as the temperature of the critical point, $T_c$, are also given. The lowest AAD values for each fluid are in bold.

| Substances | AAD | | | $T_o$ (K) | $T_f$ (K) | $T_c$ (K) |
|---|---|---|---|---|---|---|
| | **MP** | **ZTM3** | **ZTM4** | | | |
| REFRIGERANTS | | | | | | |
| R13 | 7.14 | 10.3 | **5.23** | 92.00 | 300.95 | 302 |
| R14 | 2.63 | 5.07 | **1.68** | 98.94 | 226.87 | 227.51 |
| R32 | 4.72 | 5.77 | **1.70** | 136.34 | 350.18 | 351.255 |
| R23 | 4.41 | 7.99 | **3.55** | 118.02 | 298.39 | 299.293 |
| R41 | 1.75 | 4.64 | **1.15** | 175.00 | 316.57 | 317.28 |
| R123 | 8.35 | 7.59 | **2.76** | 166.00 | 455.38 | 456.831 |
| R125 | 2.98 | 5.74 | **1.57** | 172.52 | 338.34 | 339.173 |
| R134a | 3.41 | 8.11 | **3.68** | 169.85 | 373.19 | 374.21 |
| R141b | 7.78 | 9.21 | **4.34** | 169.68 | 475.96 | 477.5 |
| R142b | 6.95 | 8.55 | **4.00** | 142.72 | 408.92 | 410.26 |
| R143a | 2.70 | 6.26 | **2.29** | 161.34 | 344.93 | 345.857 |
| R152a | 5.82 | 7.05 | **2.70** | 154.56 | 385.25 | 386.411 |
| R218 | 9.50 | 10.2 | **4.78** | 125.45 | 343.92 | 345.02 |
| R227ea | **3.52** | 8.82 | 3.71 | 146.35 | 374.80 | 375.95 |
| RC318 | 2.25 | 5.17 | **0.96** | 233.35 | 387.60 | 388.38 |
| HYDROCARBONS | | | | | | |
| Butane | 5.35 | 8.72 | **4.07** | 134.90 | 423.67 | 425.125 |
| Decane | **4.22** | 8.77 | 4.27 | 243.50 | 615.83 | 617.7 |
| Dodecane | **1.65** | 13.5 | 7.45 | 263.60 | 656.13 | 658.1 |
| Ethane | 4.55 | 8.22 | **4.36** | 90.352 | 304.26 | 305.33 |
| Ethene | **1.82** | 7.18 | 3.53 | 103.99 | 281.46 | 282.35 |
| Heptane | 6.83 | 9.28 | **4.60** | 182.55 | 538.34 | 540.13 |
| Hexane | 5.94 | 7.97 | **3.58** | 177.83 | 506.17 | 507.82 |
| Isobutane | 12.5 | 10.4 | **5.45** | 113.73 | 406.34 | 407.81 |
| Methane | 2.03 | 4.71 | **1.09** | 90.694 | 190.06 | 190.564 |
| Nonane | **1.65** | 12.4 | 7.54 | 219.70 | 592.68 | 594.55 |
| Octane | **4.09** | 9.07 | 4.68 | 216.37 | 567.56 | 569.32 |
| Pentane | **3.60** | 11.9 | 7.72 | 143.47 | 468.07 | 469.7 |
| Propane | 11.2 | 12.9 | **8.15** | 85.48 | 368.40 | 369.825 |
| | | | OTHERS | | | |
| Argon | 2.17 | 3.94 | **0.65** | 83.806 | 150.35 | 150.687 |
| Carbon dioxide | 1.83 | 2.90 | **0.09** | 216.59 | 303.69 | 304.1282 |
| Carbon monoxide | **0.68** | 4.70 | 2.49 | 68.16 | 132.54 | 132.86 |
| Deuterium oxide | 10.8 | 6.24 | **2.97** | 276.97 | 642.06 | 643.89 |
| Hydrogen | 1.67 | 2.90 | **1.07** | 13.957 | 33.049 | 33.145 |
| Hydrogen sulfide | 2.92 | 3.48 | **0.10** | 187.70 | 372.17 | 373.1 |
| Krypton | 1.45 | 3.93 | **1.12** | 115.77 | 209.01 | 209.48 |
| Nitrogen | 2.58 | 4.67 | **1.34** | 63.151 | 125.88 | 126.192 |
| Oxygen | 6.73 | 4.50 | **1.47** | 54.361 | 154.08 | 154.581 |
| Parahydrogen | 1.64 | 3.01 | **1.14** | 13.80 | 32.842 | 32.938 |
| Water | 12.5 | 5.29 | **3.40** | 273.16 | 645.23 | 647.096 |
| Xenon | 1.81 | 3.54 | **0.26** | 161.40 | 289.09 | 289.733 |

**Table 2.** Number of fluids satisfying different AAD values ranges when using the MP, ZTM3, and ZTM4 correlations, which are Eqs. (3), (8), and (9), respectively.

| AAD range | Number of fluids | | |
|---|---|---|---|
| | **MP** | **ZTM3** | **ZTM4** |
| <1% | 1 | 0 | 5 |
| <5% | 26 | 12 | 34 |
| <10% | 36 | 32 | 40 |
| >10% | 4 | 8 | 0 |

**Table 3.** Coefficients for the MP correlation, Eq. (3).

| Substances | $A_1$ | $B_1$ | $\phi$ |
|---|---|---|---|
| REFRIGERANTS | | | |
| R13 | -13.3915 | -3.9189 | 1.0085 |
| R14 | -40.1532 | -10.7951 | 1.3427 |
| R23 | -17.5927 | -5.3212 | 1.1111 |
| R32 | -20.8921 | -6.6458 | 1.1726 |
| R41 | -27.9232 | -7.9999 | 1.2306 |
| R123 | -11.1483 | -3.2854 | 1.0186 |
| R125 | -14.7809 | -4.0703 | 1.0645 |
| R134a | -14.9727 | -4.2985 | 1.0689 |
| R141b | -12.1469 | -3.6762 | 0.9914 |
| R142b | -13.9707 | -4.1386 | 1.0310 |
| R143a | -22.9821 | -6.4599 | 1.1526 |
| R152a | -18.9187 | -5.5870 | 1.0756 |
| R218 | -8.6844 | -2.3697 | 0.8752 |
| R227ea | -9.0870 | -2.5588 | 0.8966 |
| RC318 | -11.0994 | -2.9079 | 1.0124 |
| HYDROCARBONS | | | |
| Butane | -21.1764 | -6.3322 | 1.0486 |
| Decane | -18.2083 | -5.2024 | 1.0600 |
| Dodecane | -10.9456 | -3.1727 | 0.8515 |
| Ethane | -42.6668 | -12.5193 | 1.2063 |
| Ethene | -45.8672 | -13.2451 | 1.2523 |
| Heptane | -18.4369 | -5.4569 | 1.0353 |
| Hexane | -22.4724 | -6.6054 | 1.1024 |
| Isobutane | -17.7318 | -5.2306 | 0.9968 |
| Methane | -114.5597 | -30.7434 | 1.3308 |
| Nonane | -15.4134 | -4.4666 | 0.9755 |
| Octane | -20.0316 | -5.7694 | 1.0544 |
| Pentane | -22.7869 | -6.7330 | 1.0864 |
| Propane | -19.6038 | -5.9981 | 1.0176 |
| OTHERS | | | |
| Argon | -61.7837 | -15.4106 | 1.3406 |
| Carbon dioxide | -63.6491 | -17.3824 | 1.4761 |
| Carbon monoxide | -216.8303 | -51.6697 | 1.7986 |
| Deuterium oxide | -121.0082 | -46.7489 | 1.8659 |
| Hydrogen | -3.8402e+004 | -6.7737e+003 | 1.9019 |
| Hydrogen sulfide | -21.1048 | -6.9561 | 1.1721 |
| Krypton | -44.6862 | -11.5761 | 1.4181 |
| Nitrogen | -112.9249 | -26.6328 | 1.3602 |
| Oxygen | -87.6122 | -23.3290 | 1.5266 |
| Parahydrogen | -3.2494e+004 | -5.7321e+003 | 1.8678 |
| Water | -202.6043 | -78.5458 | 2.0143 |
| Xenon | -17.6010 | -5.0054 | 1.2502 |

**Table 4.** Coefficients for the ZTM3 correlation, Eq. (8).

| Substances | $A_2$ | $B_2$ | $C_2$ |
|---|---|---|---|
| REFRIGERANTS | | | |
| R13 | -10.0048 | -0.6820 | -0.0919 |
| R14 | -105.0468 | -99.6598 | -0.9740 |
| R23 | -13.6179 | -1.6785 | -0.1526 |
| R32 | -74.0657 | -73.9380 | -1.0364 |
| R41 | 4.1941 | -0.3933 | 0.0561 |
| R123 | -15.0164 | -1.7734 | -0.1407 |
| R125 | 12.0908 | -1.4749 | 0.1005 |
| R134a | -28.4553 | -6.8104 | -0.2629 |
| R141b | -10.3897 | -0.8285 | -0.1055 |
| R142b | -10.8487 | -0.8781 | -0.1059 |
| R143a | 34.3514 | -12.2087 | 0.3290 |
| R152a | -16.7765 | -2.3332 | -0.1665 |
| R218 | -9.8271 | -0.4861 | -0.0650 |
| R227ea | -77.1721 | -42.6296 | -0.5707 |
| RC318 | 5.1924 | -0.2662 | 0.0349 |
| HYDROCARBONS | | | |
| Butane | -13.3620 | -1.3864 | -0.1330 |
| Decane | -12.6904 | -1.1220 | -0.1124 |
| Dodecane | 75.1852 | -45.6301 | 0.5844 |
| Ethane | -9.7158 | -0.6553 | -0.0989 |
| Ethene | -28.9079 | -8.1476 | -0.3153 |
| Heptane | -8.5601 | -0.4743 | -0.0820 |
| Hexane | -9.5187 | -0.6268 | -0.0933 |
| Isobutane | -7.1072 | -0.2719 | -0.0635 |
| Methane | 2.7386 | -0.2688 | 0.0422 |
| Nonane | -12.4291 | -1.0551 | -0.1092 |
| Octane | -12.1214 | -1.0188 | -0.1094 |
| Pentane | -7.0539 | -0.3021 | -0.0708 |
| Propane | -6.3115 | -0.2240 | -0.0649 |
|              |  OTHERS |         |         |
| Argon            | 2.1871   | -0.1323 | 0.0250  |
| Carbon dioxide   | 0.7030   | -0.0853 | 0.0235  |
| Carbon monoxide  | 11.9948  | -1.6323 | 0.1095  |
| Deuterium oxide  | -3.6007  | -0.1020 | -0.0994 |
| Hydrogen         | -21.8828 | -0.5778 | -0.0348 |
| Hydrogen sulfide | 2.4490   | -0.2954 | 0.0554  |
| Krypton          | 6.3553   | -0.5866 | 0.0650  |
| Nitrogen         | 19.9277  | -2.8182 | 0.1209  |
| Oxygen           | -6.6956  | -0.2122 | -0.0550 |
| Parahydrogen     | -28.0718 | -1.0479 | -0.0460 |
| Water            | -3.5915  | -0.0929 | -0.0985 |
| Xenon            | 1.5134   | -0.0967 | 0.0243  |

**Table 5.** Coefficients for the ZTM4 correlation, Eq. (9).

| Substances | $A_3$ | $B_3$ | $C_3$ | $1/n$ |
|---|---|---|---|---|
| REFRIGERANTS | | | | |
| R13    | -3.3574 | 0.0176 | -1.0148 | -0.0053 |
| R14    | -3.4191 | 0.0016 | -1.0014 | -0.0005 |
| R23    | -3.3769 | 0.0115 | -1.0090 | -0.0035 |
| R32    | -3.5456 | 0.0034 | -1.0020 | -0.0009 |
| R41    | -3.7087 | 0.0134 | -1.0061 | -0.0028 |
| R123   | -3.2919 | 0.0053 | -1.0037 | -0.0014 |
| R125   | -3.5689 | 0.0195 | -1.0097 | -0.0039 |
| R134a  | -3.4093 | 0.0061 | -1.0042 | -0.0016 |
| R141b  | -3.3561 | 0.0142 | -1.0112 | -0.0042 |
| R142b  | -3.3601 | 0.0056 | -1.0046 | -0.0017 |
| R143a  | -3.6168 | 0.0053 | -1.0033 | -0.0013 |
| R152a  | -3.5963 | 0.0036 | -1.0026 | -0.0010 |
| R218   | -3.2426 | 0.0054 | -1.0046 | -0.0015 |
| R227ea | -3.5219 | 0.0120 | -1.0066 | -0.0025 |
| RC318  | -3.5346 | 0.0078 | -1.0032 | -0.0013 |
| HYDROCARBONS | | | | |
| Butane    | -3.7439 | 0.0170 | -1.0129 | -0.0049 |
| Decane    | -3.4899 | 0.0046 | -1.0039 | -0.0014 |
| Dodecane  | -4.0434 | 0.0088 | -1.0039 | -0.0016 |
| Ethane    | -3.7713 | 0.0064 | -1.0063 | -0.0023 |
| Ethene    | -3.8008 | 0.0039 | -1.0031 | -0.0012 |
| Heptane   | -3.5038 | 0.0014 | -1.0014 | -0.0005 |
| Hexane    | -3.5095 | 0.0135 | -1.0128 | -0.0046 |
| Isobutane | -3.5300 | 0.0080 | -1.0088 | -0.0030 |
| Methane   | -4.4563 | 0.0153 | -1.0082 | -0.0034 |

| Nonane | -3.6095 | 0.0128 | -1.0099 | -0.0036 |
| Octane | -3.5988 | 0.0100 | -1.0083 | -0.0030 |
| Pentane | -3.4450 | 0.0091 | -1.0103 | -0.0036 |
| Propane | -3.5015 | 0.0090 | -1.0109 | -0.0038 |
| OTHERS | | | | |
| Argon | -3.9705 | 0.0136 | -1.0072 | -0.0028 |
| Carbon dioxide | -3.5576 | 0.5273 | -1.2000 | -0.0897 |
| Carbon monoxide | -3.4819 | 0.0027 | -1.0031 | -0.0010 |
| Deuterium oxide | -2.7930 | 0.0074 | -1.0209 | -0.0088 |
| Hydrogen | -5.7707 | 0.0083 | -1.0183 | -0.0038 |
| Hydrogen sulfide | -3.5976 | 0.9351 | -1.2093 | -0.1287 |
| Krypton | -3.4083 | 0.0094 | -1.0067 | -0.0025 |
| Nitrogen | -4.1025 | 0.0214 | -1.0194 | -0.0060 |
| Oxygen | -3.3419 | 0.0101 | -1.0160 | -0.0050 |
| Parahydrogen | -5.7983 | 0.0075 | -1.0158 | -0.0033 |
| Water | -2.9036 | 0.0016 | -1.0047 | -0.0020 |
| Xenon | -3.4403 | 0.0117 | -1.0037 | -0.0020 |

**Table 6.** Comparison between the values of the viscosity at the critical point from the ZTM4 correlation, $\eta_c |_{ZTM4}$, and from the NIST database, $\eta_c |_{NIST}$. PD is the percentage deviation. The results for 21 substances are listed. For the other 19 substances, the critical viscosity is not given in the NIST database.

| Substances | $\eta_c |_{NIST}$ | $\eta_c |_{ZTM4}$ | PD (%) for *ZTM4* |
|---|---|---|---|
| R14 | 0.034206 | 0.0327 | 4.40 |
| R41 | 0.031815 | 0.0242 | 23.94 |
| R134a | 0.034686 | 0.0329 | 5.15 |
| R142b | 0.033984 | 0.0345 | -1.52 |
| R227ea | 0.033585 | 0.0292 | 13.06 |
| Decane | 0.02932 | 0.0304 | -3.68 |
| Dodecane | 0.018929 | 0.0174 | 8.08 |
| Ethane | 0.02184 | 0.0229 | -4.86 |
| Ethene | 0.022883 | 0.0223 | 2.55 |
| Heptane | 0.026644 | 0.0300 | -12.60 |
| Hexane | 0.02645 | 0.0295 | -11.53 |
| Isobutane | 0.024421 | 0.0291 | -19.16 |
| Nonane | 0.023586 | 0.0267 | -13.20 |
| Octane | 0.028789 | 0.0271 | 5.87 |
| Pentane | 0.024137 | 0.0316 | -30.92 |
| Carbon monoxide | 0.032117 | 0.0307 | 4.41 |
| Deuterium oxide | 0.038794 | 0.0608 | -56.73 |
| Hydrogen | 0.003514 | 0.0031 | 11.78 |
| Hydrogen sulfide | 0.033847 | 0.0126 | 62.77 |
| Krypton | 0.040262 | 0.0328 | 18.53 |
| Parahydrogen | 0.003515 | 0.0030 | 14.65 |